\documentclass[english,twocolumn,superscriptaddress]{revtex4-2}

\usepackage{newtxmath}
\usepackage[T1]{fontenc}
\usepackage[latin9]{inputenc}
\usepackage{graphicx}
\usepackage{esint}

\makeatletter
\usepackage{newtxtext}
\usepackage{newtxmath}
\usepackage{bbm}
\usepackage{hyperref}
\usepackage{xcolor}
\definecolor{MYGreen}{RGB}{0, 100, 0}
\hypersetup{
    colorlinks=true,
    linkcolor=MYGreen,
    citecolor=MYGreen,
    urlcolor=MYGreen
}

\makeatother

\usepackage{babel}
\begin{document}
\title{Nearly semi-elliptic relation between the minimal conductivity and
Hall conductivity in unpaired Dirac fermions}
\author{Bo Fu}
\email{fubo@gbu.edu.cn}

\address{School of Sciences, Great Bay University, Dongguan 523000, China}
\author{Kai-Zhi Bai}
\address{Department of Physics, The University of Hong Kong, Pokfulam Road,
Hong Kong, China}
\author{Shi-Hao Bi}
\address{Department of Physics, The University of Hong Kong, Pokfulam Road,
Hong Kong, China}
\author{Shun-Qing Shen}
\email{sshen@hku.hk}

\address{Department of Physics and State Key Laboratory of Optical Quantum
Matter, The University of Hong Kong, Pokfulam Road, Hong Kong, China}
\date{\today}
\begin{abstract}
Electric conductivities may reveal the topological and magnetic properties
of band structures in solids, especially for two-dimensional unpaired
Dirac fermions. In this work, we evaluate the longitudinal and Hall
conductivity for unpaired Dirac fermions in the framework of the self-consistent
Born approximation and find a nearly semi-elliptic relation between
the minimal conductivity and Hall conductivities in the Dirac fermions.
Near the charge neutrality point, disorder may drive a metal-insulator
transition, and enhance the longitudinal conductivity. For the massless
case, the minimal conductivity $\sigma_{xx}^{*}$ coexists with the
half-quantized Hall conductivity $\frac{e^{2}}{2h}$, forming an indicator
for the parity anomalous semimetal. The relation signals a disorder-induced
metallic phase that bridges two topologically distinct insulating
phases, and agrees with the recent experimental observation in magnetic
topological insulators.
\end{abstract}
\maketitle

\paragraph*{Introduction.}

The interplay between topology and magnetism in electronic systems
has unveiled a rich landscape of quantum phenomena \citep{Nagaosa-10rmp,xiao2010berry,nomura2011surface,wang2015chiral,liu2016quantum,zhang2017anomalous,liu2008quantum,liu2018intrinsic,mogi2015magnetic,wang2021intrinsic,mogi2017magnetic,tokura2019magnetic,wang2022fractional,Mogi2022:NatPhys,lu2024fractional,lu2021half,ChangCZ-23rmp}.
A paramount achievement in this field is the quantum anomalous Hall
effect (QAHE), a zero-magnetic-field state characterized by a quantized
Hall conductance \citep{thouless1982quantized,Haldane1988:PRL,yu2010quantized,Chu2011:PRB,chang2013experimental,checkelsky2014trajectory,chang2015high}.
This state was first experimentally realized in magnetically doped
topological insulator (TI) films, where the introduction of ferromagnetic
order opens a gap in the Dirac surface states, revealing a topologically
non-trivial ground state. The observation of the QAHE has since spurred
intense research into its fundamental nature, particularly its connection
to the more established integer quantum Hall effect (IQHE) in a strong
magnetic field \citep{checkelsky2014trajectory,grauer2017scaling}.
A cornerstone of the IQHE theory is the universal scaling behavior
of the longitudinal ($\sigma_{xx}$) and Hall ($\sigma_{xy}$) conductivities
near critical points \citep{huckestein1995scaling,pruisken1988universal}.
Pioneering works demonstrated that the transition between IQHE plateaus
follows a semicircle law in the $\sigma_{xy}-\sigma_{xx}$ plane,
a relation that has been instrumental in analyzing critical transitions
\citep{kivelson1992global,dykhne1994theory,ruzin1995universal,dolan1999modular}.
Recent experimental advances have enabled dynamic control of magnetization
and the band gaps of the consequent surface states in a trilayer structures
of V-doped $(\mathrm{Bi,Se})_{2}\mathrm{Te}_{3}$ (V-BST)/$(\mathrm{Bi,Se})_{2}\mathrm{Te}_{3}$
(BST)/ Cr-doped $(\mathrm{Bi,Se})_{2}\mathrm{Te}_{3}$ (Cr-BST) \citep{wang2025parity,zhuo2025evidence}.
In these systems, the chemical potential can be precisely tuned to
the charge neutrality point via gate voltage, accessing the Chern
insulator, axion insulator \citep{mogi2017magnetic,zhang2019topological,mogi2017tailoring,xiao2018realization,liu2020robust},
and parity-anomalous semimetal (PAS) \citep{Fu2022:npjQM,Zou2022:PRB,Shen-24coshare}.
Intriguingly, these systems display diverse flow diagrams under in-plane
and out-of-plane magnetic fields that deviates significantly from
the conventional semicircle form. However, the fundamental origin
of this flow diagram remains an open and critical question.

\begin{figure}
\includegraphics[width=8cm]{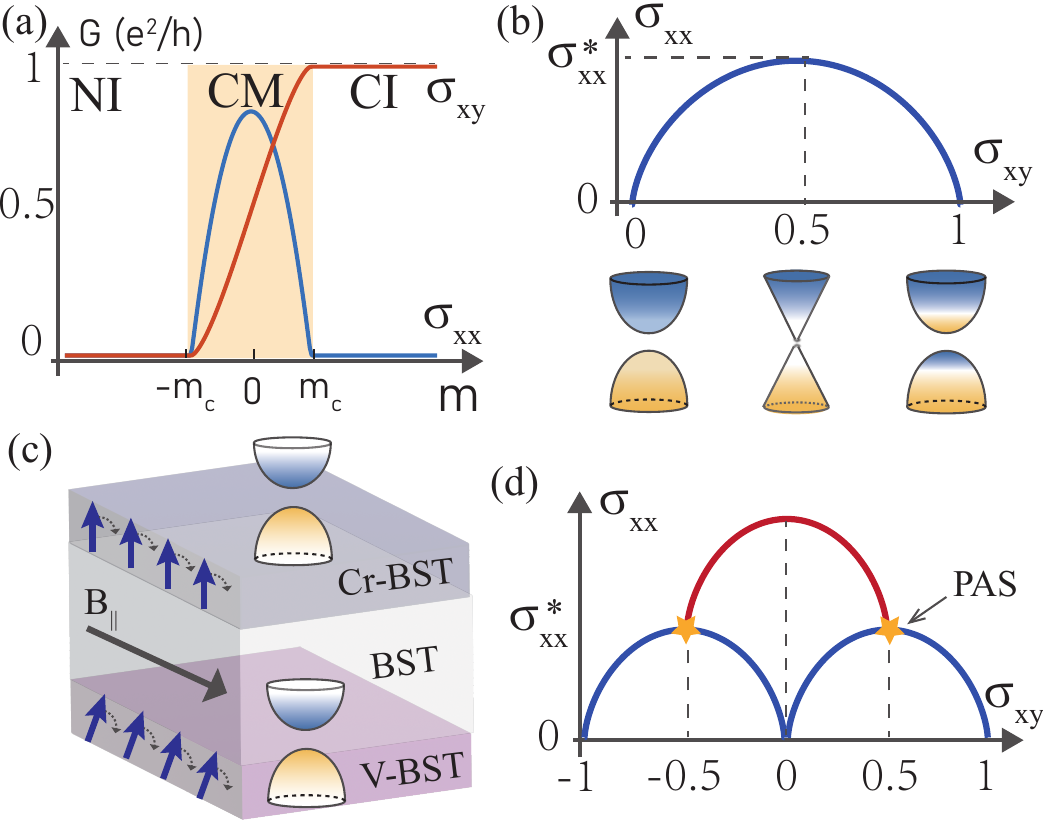}

\caption{(a) The longituidnal ($\sigma_{xx}$) and Hall ($\sigma_{xy}$) conductivities
at $E=0$ as a function of the energy gap. The yellow-shaded region
($-m_{c}<m<m_{c}$) indicates an intermediate critical metallic (CM)
phase, characterized by a non-quantized Hall conductivity and a non-zero
$\sigma_{xx}$, which separates two insulating phases with distinct
Chern numbers ($\mathcal{C}=0$ and $\mathcal{C}=1$). The critical
mass $m_{c}=2E_{c}e^{-1/2\gamma}$ depends on the disorder strength
$\gamma$. The lower panel provides a schematic of the band inversion
process, where the color gradient (blue to yellow) represents the
sign of the Berry curvature. (b) Near semi-elliptical flow diagram
of ($\sigma_{xy},\sigma_{xx}$) based on Eq. (1), resulting from the
formation of the intermediate metallic Hall phase. The peaks of the
trajectory occur at $(\sigma_{xy},\sigma_{xx})=(\pm e^{2}/2h,\sigma_{xx}^{*})$
which corresponds to a band gap closure. (c) Schematic of the heterostructure:
a magnetic TI sandwich, with the top and bottom layers doped with
Cr and V, respectively. The difference in coercive fields allows an
in-plane magnetic field ($B_{||}$) to independently switch the magnetic
orientation of each layer, thereby directly tuning the band gap of
the surface states. (d)Characteristic flow diagram for the magnetic
TI sandwich under $B_{||}$. The yellow pentangle indicates PAS. \protect\label{fig:illustration_diagram}}
\end{figure}

In this work, we evaluated the longitudinal and Hall conductivity
for the Dirac fermions in the framework of the self-consistent Born
approximation (SCBA), revealing that both are governed by a single
dimensionless parameter: the ratio between the mass-induced gap and
the disorder broadening at charge neutrality point. This leads to
a nearly semi-elliptic relation between $\sigma_{xx}$ and $\sigma_{xy}$
that specifically accounts for the unique properties of gapped Dirac
fermion bands (in the unit of $\frac{e^{2}}{h}$, Fig. \ref{fig:illustration_diagram}(b))

\begin{equation}
\left(\left|\left|\sigma_{xy}\right|-\frac{1}{2}\right|-\sigma_{c}\right)^{2}+\frac{1}{(\pi\sigma_{xx}^{*})^{2}}\left(\sigma_{xx}-\frac{\sigma_{xx}^{*}}{2}\right)^{2}=\frac{1}{4\pi^{2}}\label{eq:semi-elliptic}
\end{equation}
where $\sigma_{c}=\frac{1}{\pi}\arccos\sqrt{\sigma_{xx}/\sigma_{xx}^{*}}$
and its value is limited in the range of $0\leq\sigma_{c}\leq\frac{1}{2}$.
$\sigma_{xx}^{*}$ is the minimal conductivity at $\sigma_{xy}=\pm\frac{e^{2}}{2h}$
for the massless Dirac fermions or parity anomalous semimetal. The
point $(\sigma_{xx},\sigma_{xy})$ traces a trajectory along this
ellipse while the relevant parameters varies, demonstrating a universal
scaling behavior of unpaired Dirac fermions. This scaling signals
a disorder-induced critical metallic phase that bridges two topologically
distinct insulating states {[}Fig. \ref{fig:illustration_diagram}(a){]}---a
mechanism that fundamentally constrains the conductivities through
the interplay of topology and disorder. Our theory, when applied to
the TI sandwich heterostructure (Fig. \ref{fig:illustration_diagram}(c)),
quantitatively explains the characteristic two-stage anomalous flow
observed under an in-plane magnetic field (Fig. \ref{fig:illustration_diagram}(d)).
This flow provides direct evidence for a parity-anomalous semimetal
state. The model successfully reconciles theory with experiment by
accounting for the diverse flow diagrams reported in recent studies.

\paragraph*{Model Hamiltonian.}

To model the transport properties of the magnetic TI film, we construct
an effective theory by projecting the full three-dimensional Hamiltonian
onto the subspace spanned by the four lowest-energy bands related
to the surface band \citep{zou2023half,BaiKZ-24SciPost},
\begin{equation}
H_{0}=\left(\begin{array}{cc}
H_{t} & M\\
M^{\dagger} & H_{b}
\end{array}\right).
\end{equation}
Here, $H_{t/b}$ denote the Hamiltonians for the top ($t$) and bottom
($b$) surface states at low energies, 
\begin{equation}
H_{t/b}=\hbar v(k_{x}\sigma_{x}+k_{y}\sigma_{y})+m_{t/b}\sigma_{z}\label{eq:low_energy_kp}
\end{equation}
where $m_{t/b}$ are the mass terms induced by an out-of-plane Zeeman
field, and $v$ is the Fermi velocity. The off-diagonal block $M$
encodes the coupling between the two surface subblocks, which is significant
only at high energies: $M(\mathbf{k})=\Theta(b\mathbf{k}^{2}-m_{0})(m_{0}-b\mathbf{k}^{2})\sigma_{z}$.
Here $\Theta(x)$ denotes the Heaviside step function. The condition
$m_{0}b>0$ is imposed to preserve the topological nature of three
dimensional bulk. Consequently, within the momentum range $k<k_{c}=\sqrt{m_{0}/b}$,
the coupling is identically zero ($M=0$).

\begin{figure}
\includegraphics[width=8cm]{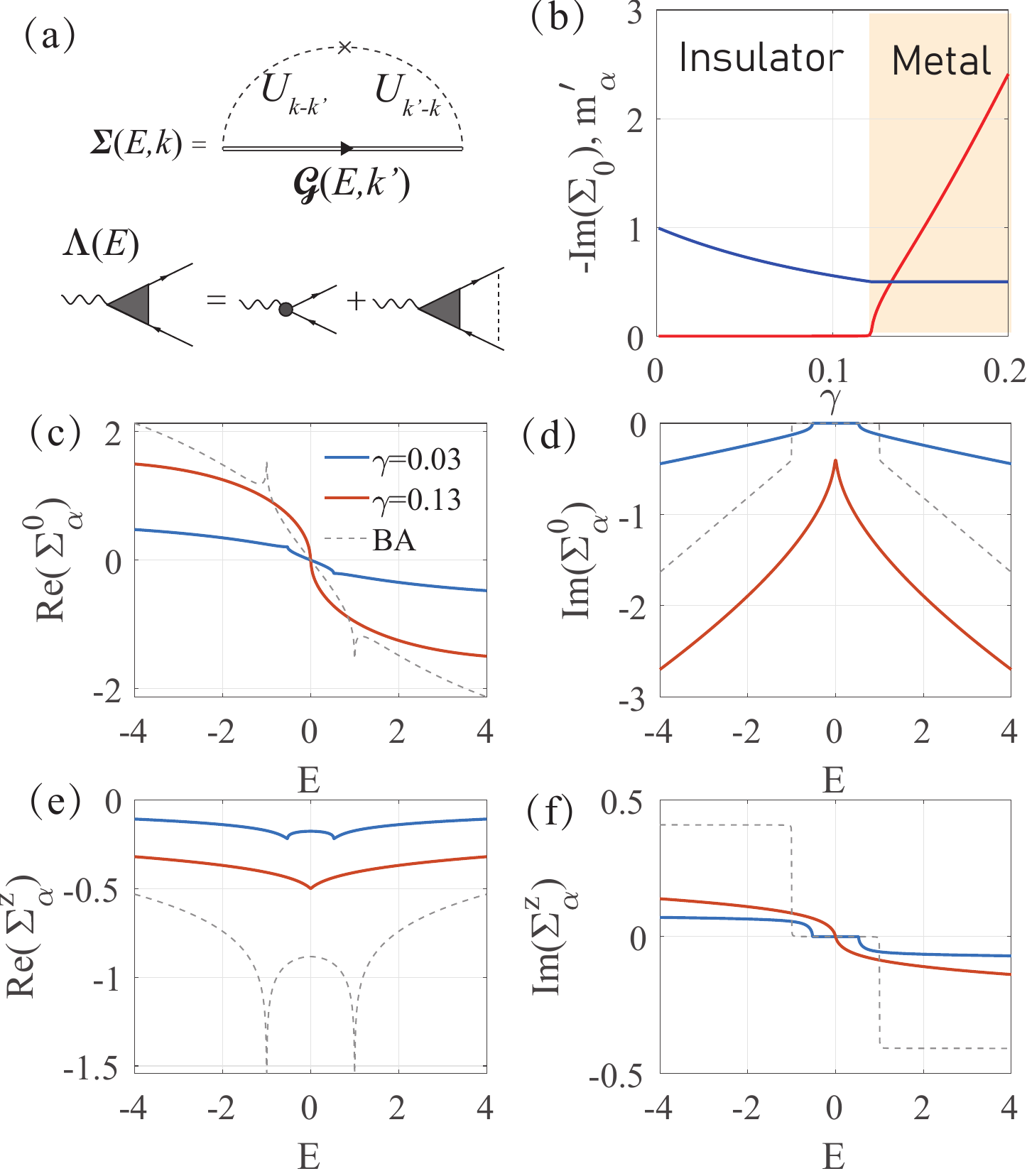}\caption{(a)Illustration of the Feynman diagram for the self-energy within
the SCBA and vertex correction. (b) Imaginary part of the self-energy
$\Sigma_{\alpha}^{0}$ (red) and the renormalized gap $m_{\alpha}^{\prime}$
(blue) as a function of disorder strength $\gamma$ at zero energy.
(c-f) Real and imaginary parts of the self-energies $\Sigma_{\alpha}^{0}$
and $\Sigma_{\alpha}^{z}$ as a function of energy $E$ for $\gamma=0.03$
(below the critical disorder $\gamma_{\alpha,c}$) and $\gamma=0.13$
(above $\gamma_{\alpha,c}$). Results from the Born approximation
(BA, dashed lines) are shown for comparison. The bare gap $m_{\alpha}=1$
is used as the energy unit.\protect\label{fig:SCBA}}
\end{figure}

\paragraph*{Self-consistent Born approximation and Insulator-Metal transition
induced by disorder.}

To systematically treat the effects of quenched disorder, we employ
the SCBA to evaluate the self energy. The disorder is modeled as a
Gaussian white noise potential with the correlator:$\langle U_{\mathbf{k}-\mathbf{k}^{\prime}}U_{\mathbf{k}^{\prime}-\mathbf{k}}\rangle_{dis}=4\pi\gamma(\hbar v)^{2}$
where $\gamma$ is the dimensionless disorder strength. Within the
SCBA framework, the self-energy $\Sigma(E,\mathbf{k})$, which renormalizes
the electron's energy and lifetime, becomes momentum-independent.
It is determined by the self-consistent equation, shown diagrammatically
in Fig. \ref{fig:SCBA}(a) \citep{mahan2000many},
\begin{equation}
\Sigma_{\alpha}^{R/A}(E)=\left\langle \int\frac{d^{2}\mathbf{k}^{\prime}}{(2\pi)^{2}}U_{\mathbf{k}-\mathbf{k}^{\prime}}\mathcal{G}_{\alpha}^{R/A}(E,\mathbf{k}^{\prime})U_{\mathbf{k}^{\prime}-\mathbf{k}}\right\rangle _{{\rm dis}},
\end{equation}
where $\alpha=t,b$ and $\left\langle \cdots\right\rangle _{{\rm dis}}$
denotes the disorder-averaging. The corresponding retarded and advanced
Green's functions for the subsystems are: $\mathcal{G}_{\alpha}^{R/A}(E,\mathbf{k})=(E-\Sigma_{\alpha}^{R/A}-H_{\alpha})^{-1}$.
The self-energy for gapped Dirac bands admits the decomposition $\Sigma_{\alpha}=\Sigma_{\alpha}^{0}\sigma_{0}+\Sigma_{\alpha}^{z}\sigma_{z}$.
Substituting the corresponding Green's function into the self-energy
equation and performing the momentum integration up to the high-energy
cut-off $E_{c}\sim\hbar vk_{c}$, we obtain the following pair of
self-consistent equations:
\begin{equation}
\Sigma_{\alpha}^{0}=-\gamma(E-\Sigma_{\alpha}^{0})\ln\left[\frac{-E_{c}^{2}}{(E-\Sigma_{\alpha}^{0})^{2}-(m_{\alpha}+\Sigma_{\alpha}^{z})^{2}}\right];
\end{equation}
\begin{equation}
\Sigma_{\alpha}^{z}=-\gamma(m_{\alpha}+\Sigma_{\alpha}^{z})\ln\left[\frac{-E_{c}^{2}}{(E-\Sigma_{\alpha}^{0})^{2}-(m_{\alpha}+\Sigma_{\alpha}^{z})^{2}}\right].
\end{equation}
This set of equations can be solved numerically for the complex self-energy
$\Sigma^{0,z}={\rm Re}\Sigma^{0,z}+{\rm i}{\rm Im}\Sigma^{0,z}$.
Here, the real part ${\rm Re}\Sigma^{0,z}$ describes energy renormalization
and band gap renormalization, respectively, and the imaginary parts
${\rm Im}\Sigma^{0,z}$ characterizes the scattering rate and quasiparticle
lifetime. At $E=0,$ the self-consistent equations simplify: $\Sigma_{\alpha}^{0}$
becomes purely imaginary, while $\Sigma_{\alpha}^{z}$ becomes purely
real. As shown in Fig. \ref{fig:SCBA}(b), $\Sigma_{\alpha}^{0}$
is initially zero for weak disorder but becomes non-zero when $\gamma$
exceeds a critical value, $\gamma_{\alpha,c}=\left(2\ln\frac{2E_{c}}{|m_{\alpha}|}\right)^{-1}$.
The retarded and advanced self-energies take the analytic form: $\Sigma_{\alpha}^{R/A}=\mp{\rm i}\mathrm{Re}\sqrt{E_{c}^{2}e^{-1/\gamma}-(m_{\alpha}/2)^{2}}$.
Concurrently, the renormalized mass parameter $m_{\alpha}^{\prime}=m_{\alpha}+\Sigma_{\alpha}^{z}$
gradually decreases to $m_{\alpha}/2$ as $\gamma$ approaches $\gamma_{\alpha,c}$
and remains constant thereafter. The density of states, given by $\nu_{\alpha}(E)=-\frac{1}{\pi}\int\frac{d^{2}\mathbf{k}}{(2\pi)^{2}}\mathrm{Im}\mathrm{Tr}[\mathcal{G}_{\alpha}^{R}(E,\mathbf{k})]$
is directly related to the imaginary part of the self-energy $\Sigma_{\alpha}^{0}$,
which can be expressed as $\nu_{\alpha}(E)=-\frac{\mathrm{Im}\Sigma_{\alpha}^{0R}}{2\pi^{2}(\hbar v)^{2}\gamma}$.
Within the SCBA, this implies that disorder can induce a phase transition
from an insulating phase (topological or trivial) to a gapless metallic
phase \citep{meyer2013disordered,bi2025half,bi2024disordered}. For
a constant disorder strength $\gamma$ , this transition occurs when
the mass parameter $m_{\alpha}$ decreases to a critical value of
$m_{\alpha,c}=2E_{c}e^{-1/2\gamma}$. In the gapless case where $m_{\alpha}=0$,
the self-energy $\Sigma_{\alpha}^{R/A}=\mp{\rm i}E_{c}e^{-1/2\gamma}$
remains finite for any infinitesimal disorder. The energy dependence
of the self-energies $\Sigma^{0}$ and $\Sigma^{z}$ is shown in Fig.
\ref{fig:SCBA}(c-f) for disorder strengths below and above $\gamma_{\alpha,c}$.
The real and imaginary parts exhibit opposite parity in energy, a
consequence of the Kramers-Kronig relations. With increasing disorder,
the band gap is gradually smeared, as evidenced by the broadening
of the features in $Im\Sigma_{\alpha}^{0}$. Furthermore, the mass
renormalization $\mathrm{Re}\Sigma_{\alpha}^{z}$ is always negative,
reducing the effective mass and thereby promoting the insulator-to-metal
transition.

\paragraph*{Charge Transport in Disordered Dirac Systems.}

The electrical conductivity tensor in the static limit can be evaluated
from the Kubo formula. The SCBA framework allows us to extract essential
physical parameters, including the renormalized energy $E_{\alpha}^{\prime}=E-\Sigma_{\alpha}^{0R}$.
Consequently, the zero-temperature conductivities for each subsystem
can be obtained as
\begin{equation}
\sigma_{xx,\alpha}=\frac{1}{2\pi}\left(1-\cos\phi_{-}^{\alpha}\left|\frac{\phi_{+}^{\alpha}}{\sin\phi_{+}^{\alpha}}\right|\right)
\end{equation}

\begin{equation}
\sigma_{xy,\alpha}=\frac{1}{2\pi}\left(\phi_{-}^{\alpha}-\frac{\sin\phi_{-}^{\alpha}}{\sin\phi_{+}^{\alpha}}\phi_{+}^{\alpha}\right)+\sigma_{xy,\alpha}^{*}
\end{equation}
where $\phi_{+}^{\alpha}=\mathrm{Arg}(m_{\alpha}^{\prime2}-E_{\alpha}^{\prime2})$,
$\phi_{-}^{\alpha}=\mathrm{Arg}(\frac{m_{\alpha}^{\prime}+E_{\alpha}^{\prime}}{m_{\alpha}^{\prime}-E_{\alpha}^{\prime}})$,
and $\sigma_{xy,\alpha}=-\frac{1}{2}\left[\mathrm{sgn}(m_{\alpha})\pm\mathrm{sgn}(b)\right]$
in which $\pm\frac{{\rm sgn}(b)}{2}$ represents the contribution
from the high-momentum regulator to ensure the cancellation of unphysical
quantized values in the total Hall conductivity when the Fermi level
lies within a gap \citep{lu2010massive}. The sign of this regulator
term depends on whether the band is topologically trivial or nontrivial.
Crucially, the regulator contributions for the two subsystems are
always opposite in sign in this sandwich system.

\begin{figure}
\includegraphics[width=8cm]{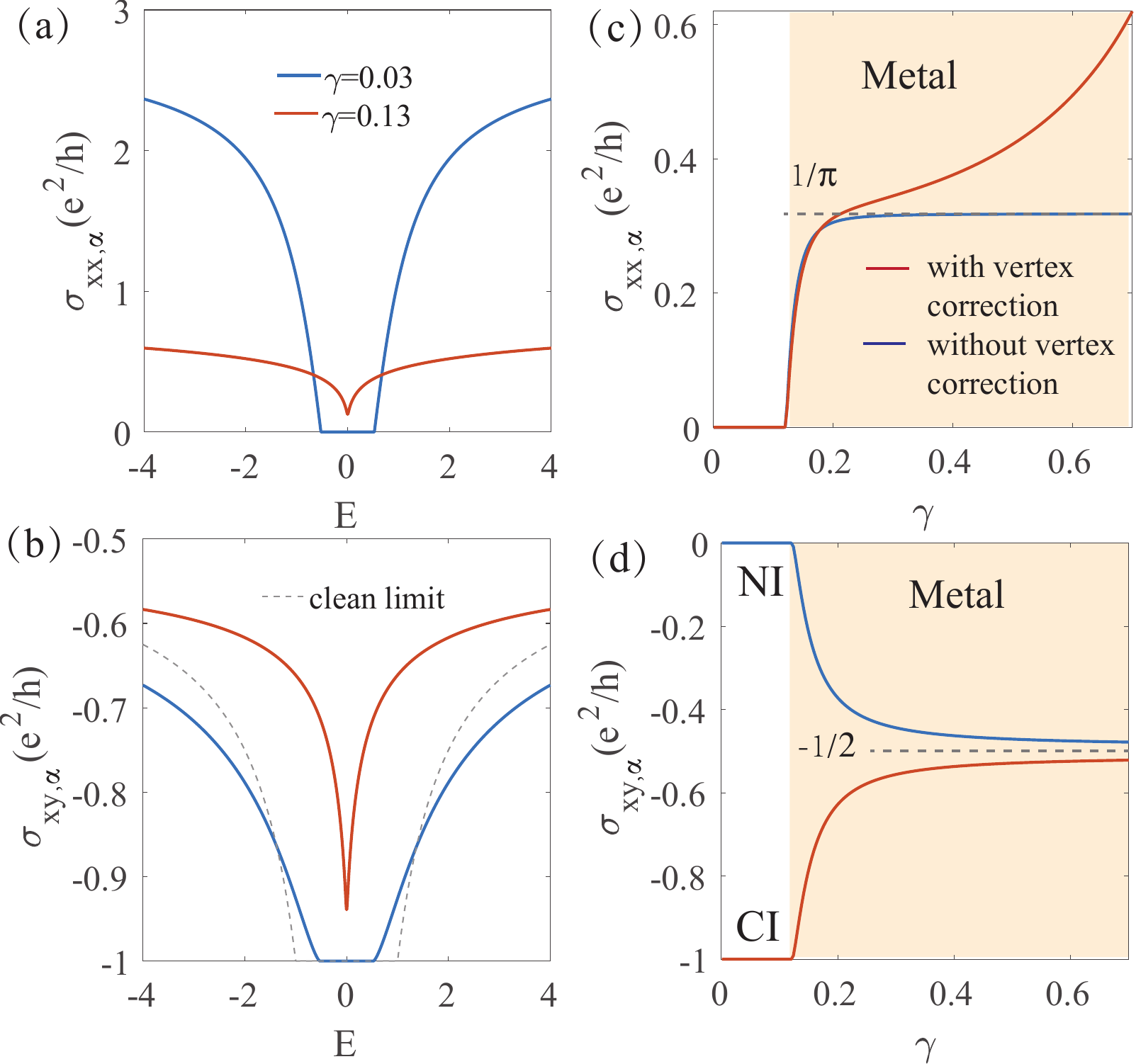}\caption{(a) Longitudinal conductivity $(\sigma_{xx,\alpha})$ and (b) Hall
conductivity $(\sigma_{xy,\alpha})$ as a function of energy for two
disorder strengths: $\gamma=0.03$ (subcritical) and $\gamma=0.13$
(supercritical). For reference, Hall conductivity in the clean-limit
($\gamma=0$) is shown as a dashed line in (b). (c) The evolution
of the zero-energy $\sigma_{xx,\alpha}$ with increasing disorder
strength $\gamma$, calculated both with (red) and without (blue)
vertex corrections. (d) Zero-energy Hall conductivity $\sigma_{xy,\alpha}$
as a function of $\gamma$ for initial states of a normal insulator
(NI, blue) and a Chern insulator (CI, red).\protect\label{fig:conductivity}}
\end{figure}

The transport behavior governed by these expressions is illustrated
in Fig. \ref{fig:conductivity}. Panel (a) shows the longitudinal
conductivity {[}$\sigma_{xx,\alpha}(E)${]}, which vanishes at the
charge neutrality point ($E=0$) below the critical disorder strength
($\gamma=0.03$), reflecting the insulating state. Once $\gamma$
exceeds $\gamma_{\alpha,c}$, a finite $\sigma_{xx,\alpha}$ emerges
at $E=0$, signaling the breakdown of the insulating phase. Panel
(b) displays the Hall conductivity {[}$\sigma_{xy,\alpha}(E)${]}.
For the subcritical disorder strength ($\gamma=0.03$), $\sigma_{xy,\alpha}$
retains the characteristic plateau structure of the clean quantum
Hall phase (dashed line) when the Fermi energy lies within the renormalized
band gap. In contrast, for a supercritical disorder strength ($\gamma=0.13$),
these plateaus are washed out and the quantization is destroyed, a
direct consequence of disorder-induced band broadening and gap closure.
This transition is summarized in panels (c) and (d), which tracks the zero-energy
($E=0$) conductivities as a function of $\gamma$. The rise of $\sigma_{xx,\alpha}$
 and the simultaneous deviation of $\sigma_{xy,\alpha}$ from its quantized value
at a critical $\gamma_{\alpha,c}$ provide direct transport signatures
of the insulator-to-metal transition induced by disorder. For large
disorder strengths, when the mass $m_{\alpha}$ is completely smeared,
the system ultimately transitions to a parity anomalous semimetal
state, characterized by $\sigma_{xx,\alpha}=\frac{e^{2}}{\pi h}$
and $\sigma_{xy,\alpha}=-\frac{e^{2}}{2h}$. Since the SCBA does not
account for localization, our theory is expected to hold when the
relevant spatial scales---such as the sample size $L$ or the temperature-dependent
phase-coherence length ---are less than the localization length.

\paragraph*{Conductivities at Charge Neutrality Point.}

For a finite chemical potential, the longitudinal 
and Hall conductivities are given by complex expressions
involving the band gap $m_{\alpha}$, the disorder broadening $\eta_{\alpha}=|\mathrm{Im}\Sigma_{0,\alpha}^{R}|$,
and the regulator parameter $b$. To isolate the universal behavior
arising from the interplay between the band gap and disorder, we focus
on the charge neutrality point ($E=0$). At this point, the longitudinal
conductivity has the minimal value as the density of states reaches
at the minimal when the chemical potential crosses the Dirac point.
When $E\to0$, one has ${\rm Im}\left(\Sigma_{\alpha}^{zR}\right)\to0$,
and $\phi_{+}\to0$, $\phi_{-}\to2\arctan\left(\eta_{\alpha}/m_{\alpha}^{\prime}\right)$.
The only relevant energy scales are the gap $m_{\alpha}^{\prime}$
and scattering induced broadening $\eta_{\alpha}$, allowing us to
define a single dimensionless parameter $\widetilde{m}_{\alpha}=m_{\alpha}^{\prime}/\eta_{\alpha}$.
$\sin\phi_{+}/\phi_{+}\rightarrow1$ and $\cos\phi_{-}=\left(\widetilde{m}_{\alpha}^{2}-1\right)/\left(1+\widetilde{m}_{\alpha}^{2}\right)$.
Thus, in this regime, the conductivity formulas simplify significantly,
\begin{equation}
\sigma_{xx,\alpha}=\sigma_{xx,\alpha}^{*}\frac{1}{1+\widetilde{m}_{\alpha}^{2}};
\end{equation}
\begin{equation}
\sigma_{xy,\alpha}=\sigma_{xy,\alpha}^{*}+\frac{1}{\pi}\left(\arctan\frac{1}{\widetilde{m}_{\alpha}}-\frac{\widetilde{m}_{\alpha}}{1+\widetilde{m}_{\alpha}^{2}}\right).
\end{equation}
In the weak scattering limit of $\widetilde{m}_{\alpha}\to\infty$,
$\sigma_{xx,\alpha}=0$ and the Hall conductivity $\sigma_{xy,\alpha}=-\frac{{\rm sgn}\left(\widetilde{m}_{\alpha}\right)\pm{\rm sgn}\left(b\right)}{2}$.
It depends on the relative sign between the Dirac mass $m_{\alpha}$
and higher energy regulator $b$. In the massless limit of $\widetilde{m}_{\alpha}\to0$,
$\sigma_{xy,\alpha}=\pm\frac{\mathrm{sgn}(b)}{2}$ and $\sigma_{xx,\alpha}=\sigma_{xx}^{*}$,
which represents the minimal longitudinal conductivity at the charge
neutrality point in the massless limit ($\widetilde{m}_{\alpha}\to0$).
When fitting experimental data, $\sigma_{xx,\alpha}^{*}$ can be treated
as a fitting parameter to account for realistic material-specific
effects. By using the trigonometric function relations, we can remove
the parameter $\widetilde{m}_{\alpha}$, and obtain the nearly semi-elliptic
equation in Eq. (\ref{eq:semi-elliptic}), the main result of the
present work. It is worth of stressing that the equation does not
contain the point $(0,\pm1/2)$, which means the Hall conductivity
is not allowed to be non-integer in the insulating case.

\paragraph*{Mechanisms for the Enhancement of $\sigma_{xx,\alpha}^{*}$.}

The measured $\sigma_{xx,\alpha}^{*}$ at the charge neutrality point
in various systems are typically around $\sim0.6\frac{e^{2}}{h}$
at 20-30mK \citep{Mogi2022:NatPhys,Yang2025Advmater,wang2025parity,zhuo2025evidence}.
The values are expected to decrease at even lower temperatures, but
are still larger than $\frac{e^{2}}{\pi h}$ calculated by means of
the SCBA \citep{ostrovsky2006electron}. This discrepancy might be
related to a long-standing debate concerning the minimal conductivity
of massless Dirac fermions in materials such as graphene and TIs.
The different limiting procedures---specifically, the order in which
the zero-frequency and zero-momentum limits are taken---yield distinct
results, such as $\frac{e^{2}}{\pi h}$, $\frac{\pi e^{2}}{8h}$,
or $\frac{\pi e^{2}}{4h}$ per valley per spin\citep{ziegler2007minimal}.
This non-universality is further complicated by scattering mechanisms.
By incorporating the vertex corrections to the current operator {[}Fig.
\ref{fig:SCBA}(a), bottom panel{]}, we find that $\sigma_{xx,\alpha}^{*}$
is modified as: $\sigma_{xx,\alpha}^{*}=\frac{e^{2}}{\pi h}\frac{1}{1-\gamma^{2}}$\citep{Fu2025SM}.
For a large scattering strength $\gamma$, this expression predicts
a significant enhancement over the bare value of $\frac{e^{2}}{\pi h}$,
as demonstrated in Fig. \ref{fig:conductivity}(c). Another possible
contribution is the warping effect of the surface states in $\mathrm{Be}_{2}\mathrm{Te}_{3}$
\citep{fu2009hexagonal,cserti2007role}. To account for the warping
effects, we incorporate a term $\lambda(k_{x}^{3}-3k_{x}k_{y}^{2})\sigma_{z}$
into $H_{t.b}$. $\sigma_{xx,\alpha}^{*}$ then depends on the dimensionless
parameter $\widetilde{\lambda}=\lambda(\mathrm{Im}\Sigma_{\alpha}^{0R})^{2}/(\hbar v)^{3}$and
is enhanced beyond its bare value as $\widetilde{\lambda}$ increases\citep{Fu2025SM}.
Furthermore, additional factors such as long-range disorder \citep{chen2023origin}
and quantum interference effects can also elevate $\sigma_{xx,\alpha}^{*}$.
A comprehensive theoretical resolution of this issue thus represents
a significant and necessary challenge for future work.

\begin{figure}
\includegraphics[width=8cm]{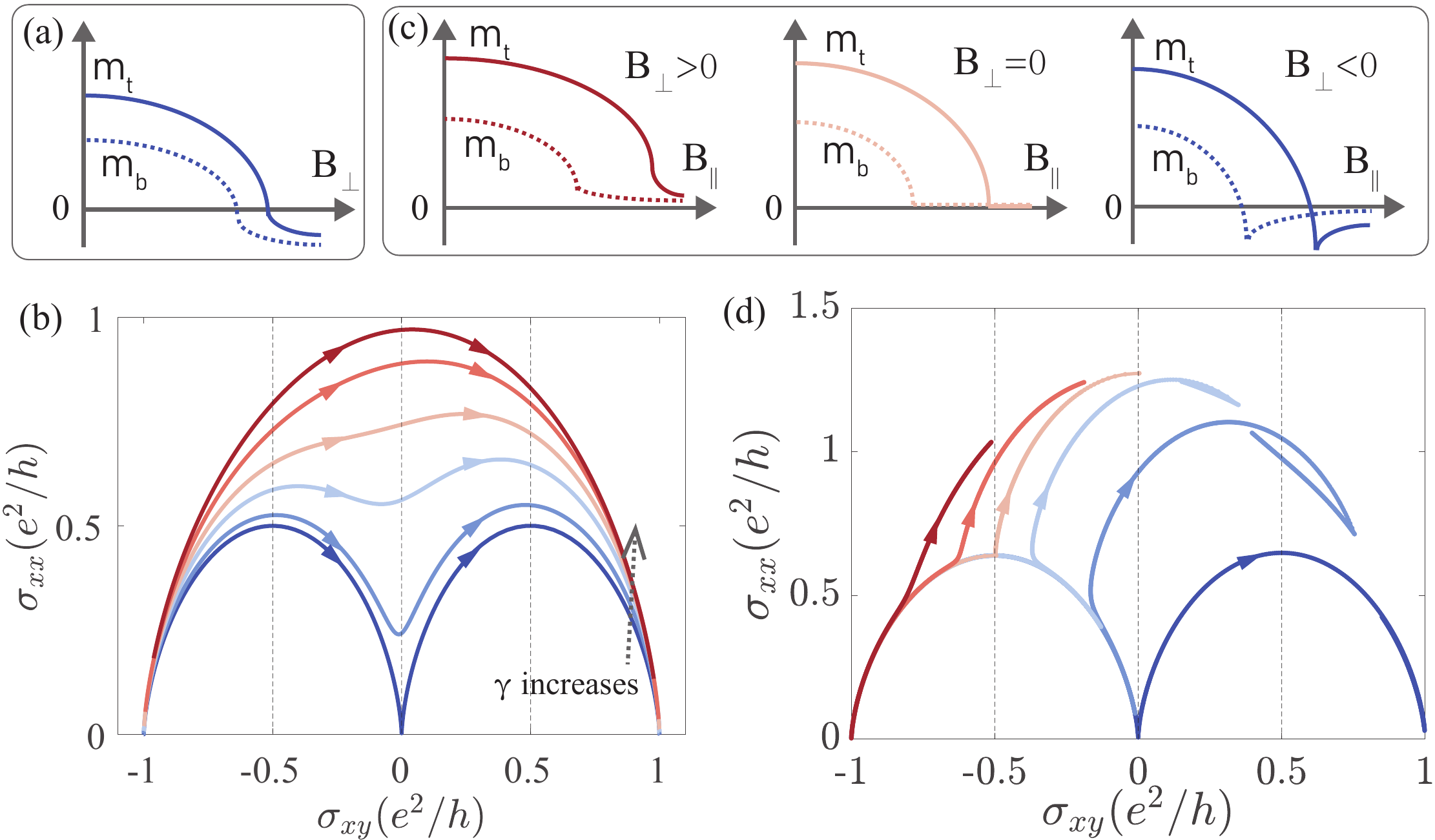}\caption{(a) Schematic of the mass terms $m_{t/b}$ in a magnetic TI sandwich
as a function of perpendicular magnetic field $B_{\perp}$. (b) Corresponding
theoretical flow diagrams in  $(\sigma_{xy},\sigma_{xx})$ plane.
The color lines from blue to red traces the evolution from the weak
to the strong scattering limit with increasing scattering strength
$\gamma$. (c)Mass terms $m_{t/b}$ under an in-plane field $B_{||}$
for different fixed values of $B_{\perp}$: $B_{\perp}>0$ (red),
$=0$ (orange), and $<0$(blue). (d) Resulting flow diagram under
the application of $B_{\parallel}$.\protect\label{fig:flow_diagram}}
\end{figure}

\paragraph*{Diverse Flow Diagrams in Trilayer systems.}

Consider the surface bands on distinct top and bottom surfaces ($\alpha=t,b$).
The total Hall and longitudinal conductivities are the sum of contributions
from both sectors. The resulting flow diagram of ($\sigma_{xy}$,$\sigma_{xx}$)
in the parameter space depends critically on the external magnetic
field, as summarized in Fig. \ref{fig:flow_diagram}. Using the Stoner-Wohlfarth
model \citep{tannous2008stoner}, we compute the mass terms $m_{t/b}$,
repsectively \citep{Fu2025SM}. Fig. \ref{fig:flow_diagram}(a) illustrates
how the Dirac mass on each surface varies with a perpendicular magnetic
field ($B_{\perp}$) and in-plane magnetic field. We consider a scenario
where the top and bottom surface masses change sign at two distinct
critical fields $B_{\perp}$, creating an intermediate phase where
the masses have opposite signs. The corresponding conductivity flow
diagrams in Fig. \ref{fig:flow_diagram}(b) with the color gradient
from blue to red traces the evolution from the weak to the strong
scattering limit with increasing disorder strength $\gamma$. In the
weak scattering limit (blue curve), where the disorder broadening
is much smaller than the individual mass gaps, the trajectory of the
flow forms a path comprised of two interconnected semi-elliptic segments
that connect at the origin $(\sigma_{xy},\sigma_{xx})=(0,0)$. In
the strong scattering regime (red curve), significant disorder broadening
smears out the intermediate axion insulator state, whose gap is relatively
small. Consequently, the flow diagram collapses into a single, semi-elliptic
curve connecting the two insulating states (e.g., from $\sigma_{xy}=+\frac{e^{2}}{h}$
to $\sigma_{xy}=-\frac{e^{2}}{h}$), with a single peak in $\sigma_{xx}$
at the gap-closing point. At intermediate disorder strengths, the
flow diagram exhibits a crossover between these two limiting behaviors.
This framework provides a natural explanation for the measured flow
diagrams in samples of varying thickness \citep{zhang2024interlayer,Fu2025SM}.

Figure \ref{fig:flow_diagram}(c) illustrates the evolution of the
mass terms under an in-plane field $B_{||}$ for different fixed values
of $B_{\perp}$. The corresponding conductivity flow diagrams are
shown in Fig. \ref{fig:flow_diagram}(d), with trajectories color-coded
from red to blue as the fixed $B_{\perp}$ value changes from positive
to negative. For $B_{\perp}=0$ , the distinct coercive fields cause
the surface Zeeman fields to orient oppositely under $B_{||}$, rendering
one surface gapless ($m_{b}=0$) while the other stays gapped $(m_{t}\ne0)$.
The resulting conductivity flow {[}orange line in Fig. \ref{fig:flow_diagram}(d){]}
exhibits a distinct two-stage evolution: in the first stage, the system
flows from its initial normal insulating state toward a PAS state
characterized by a finite longitudinal conductivity and a half-quantized
Hall conductivity; in the second stage, the net conductivity then
follows an anomalous flow trajectory, which can be understood as the
superposition of the conventional flow for a massive Dirac band and
the constant background contribution from the established parity anomalous
semimetal state. For fixed $B_{\perp}>0$ (red line), both surfaces
remain gapped, and the broken parity symmetry prevents the flow from
reaching the half-quantized $\sigma_{xy}$ point. For $B_{\perp}<0$
(blue line), the $B_{||}$-induced mass inversion leads to a flow
resembling that of varying $B_{\perp}$. At large $B_{||}$, both
Zeeman fields are forced nearly in-plane, driving the system to a
Dirac semimetal phase with $(\sigma_{xy},\sigma_{xx})=(0,2\sigma_{xx}^{*})$.
Our theory provides excellent agreement with the experimental data
obtained under in-plane magnetic fields \citep{wang2025parity,Fu2025SM}.

\paragraph*{Discussion and Summary.}

The universal nearly semi-elliptic flow diagram discovered here for
disordered massive Dirac fermions is fundamentally distinct from the
semicircle law of the IQHE. In our case, the fixed point at ($\sigma_{xy},\sigma_{xx})=(\pm\frac{e^{2}}{2h},\sigma_{xx}^{*})$
represents a PAS, where the half-quantized Hall conductivity is an
intrinsic property of the gapless Dirac cone, arising from the topological
Berry phase and coexisting with a finite minimal longitudinal conductivity.
This is a direct manifestation of pristine Dirac physics in the presence
of disorder. In stark contrast, the $\sigma_{xy}=\frac{e^{2}}{2h}$
point in the IQHE theory is not a stable phase but a critical point
of the localization transition, where a finite $\sigma_{xx}$ arises
solely from extended states inside the broadened Landau levels.

In this work, we discover a universal scaling law for topological
transport: the longitudinal and Hall conductivities of disordered
Dirac fermions at the charge neutrality point are constrained to a
nearly semi-elliptic relation. This scaling trajectory, dictated by
the interplay of topology and disorder, reveals that transitions between
distinct insulating states are necessarily mediated by an intermediate
metallic Hall phase. At the heart of this metallic phase lies the
PAS, uniquely identified by the striking coexistence of the minimal
longitudinal conductivity with the half-quantized Hall conductivity.
This universal scaling framework quantitatively explains the enigmatic
flow diagrams observed in various samples of magnetic TI sandwiches,
thereby bridging a fundamental gap between theory and experiment in
the landscape of topological quantum matter.
\begin{acknowledgments}
This work was supported by the Quantum Science Center of Guangdong-Hong
Kong-Macao Greater Bay Area (Grant No. GDZX230005) and the Research
Grants Council, University Grants Committee, Hong Kong (Grants No.
C7012-21G). B.F. is financially supported by National Natural Science
Foundation of China (Grants No.12504049), Guangdong Province Introduced
Innovative R\&D Team Program (Grant No. 2023QN10X136), Guangdong Basic
and Applied Basic Research Foundation (No. 2024A1515010430 and 2023A1515140008).
\end{acknowledgments}

\bibliographystyle{bpsrev}
\bibliography{ref-ellipse}

\end{document}